\documentclass[aps,prb,twocolumn,showpacs,floatfix,superscriptaddress,amssymb,amsmath,preprintnumbers]{revtex4}
\usepackage{bm}
\usepackage{graphicx}
\usepackage{graphics}
\usepackage{amsmath}
\usepackage{amsfonts}
\usepackage{amssymb}
\usepackage{color}

\begin{document}

\newcommand{\be}   {\begin{equation}}
\newcommand{\ee}   {\end{equation}}
\newcommand{\ba}   {\begin{eqnarray}}
\newcommand{\ea}   {\end{eqnarray}}
\newcommand{\ve}  {\varepsilon}
\newcommand{\Dis} {\mbox{\scriptsize dis}}

\newcommand{\state} {\mbox{\scriptsize state}}
\newcommand{\band} {\mbox{\scriptsize band}}
\title{Disorder-mediated Kondo effect in graphene}

\author{V.\ G.\ Miranda}
\affiliation{Instituto de F\'{\i}sica, Universidade Federal Fluminense, 24210-346 Niter\'oi, RJ, Brazil}
\author{Luis G.\ G.\ V.\ Dias da Silva}
\affiliation{Instituto de F\'{\i}sica, Universidade de S\~{a}o Paulo,
C.P.\ 66318, 05315--970 S\~{a}o Paulo, SP, Brazil}
\author{C.\ H.\ Lewenkopf}
\affiliation{Instituto de F\'{\i}sica, Universidade Federal Fluminense, 24210-346 Niter\'oi, RJ, Brazil}

\date{\today}
\begin{abstract}
We study the emergence of strongly correlated states and Kondo physics in
disordered graphene. Diluted short range disorder gives rise to localized
midgap states at the vicinity of the system charge neutrality point. We show
that long-range disorder, ubiquitous in graphene,  allows for the coupling of
these localized states to an effective (disorder averaged) metallic band. The
system is described by an Anderson-like model.  We use the numerical
renormalization group method to study the distributions of Kondo temperatures
$P(T_K)$. The results show that  disorder  can lead to long logarithmic tails
in $P(T_K)$, consistent with a quantum Griffiths phase.

\end{abstract}


\maketitle


The investigation of magnetic properties in graphene has triggered intense
research activity.\cite{Uchoa:Phys.Rev.Lett.:26805:2008,Yazyev2010} The
formation of local magnetic moments has been observed by experiments on
graphene nanoribbon edges, \cite{Tao2011} hydrogenated \cite{McCreary12} and
irradiated\cite{Chen11,Nair12,McCreary12} graphene flakes. Low temperature
experiments on irradiated samples\cite{Chen11,Nair12,McCreary12} give quite
puzzling results. For low irradiation, Ref.~\onlinecite{Chen11} reports
fingerprints of the Kondo effect in the resistivity. The reported Kondo
temperature, obtained from the single-parameter scaling characteristic of
conventional $S=1/2$ Kondo systems,\cite{Costi1994,Costi00} is rather high,
$T_K \approx 10 \cdots 100$ K, with a weak dependence on the gate voltage, both
for $p$ and $n$-doping. This is at odds with the theoretical analysis,
\cite{Vojta10} that predicts an exponential dependence of $T_K$ with the
chemical potential for $n$ doping and vanishing small Kondo effect for $p$
doping. Other experiments on irradiated graphene,\cite{Nair12,McCreary12}
observed a paramagnetic susceptibility consistent with $S=1/2$ magnetic local
moments, without evidence of Kondo quenching, even at temperatures as low as 2
K.\cite{Nair12}

The Kondo effect in graphene also poses new interesting theoretical questions.
\cite{Sengupta08,Cornaglia09,Vojta10,Uchoa:Phys.Rev.Lett.:16801:2011,Fritz12}
The linear energy dependence of the graphene density of states and the
occurrence of localized states are a physical realization of a pseudogap Kondo
model,  which is known to show a rich variety of quantum critical behavior as a
function of the gate-controlled chemical potential. \cite{Vojta10,Fritz12} What
has been overlooked so far, is that disorder, ubiquitous in graphene, modifies
this picture dramatically.

In this Rapid Communication, we present a systematic study of the Kondo effect
in disordered graphene using the numerical renormalization group (NRG) method.
Disorder provides a simple coupling mechanism leading to low-temperature Kondo
screening. We find that the resulting distribution of Kondo temperatures
$P(T_K)$ depends on the disorder strength and, in a more subtle manner, on the
chemical potential.  Interestingly, we show that, as the system enters the
Kondo regime, long range disorder can lead to logarithmic tails in $P(T_K)$,
which are characteristic of a quantum Griffiths phase.
\cite{Neto:PhysicalReviewLetters:3531:1998,Miranda:Phys.Rev.Lett.:264:2001,Mirandareview}
This scenario is much richer than the standard one in dirty metals, where the
disorder is responsible mainly for a local modification of the band-impurity
coupling constant.\cite{Zarand96,Lewenkopf05,Zhuravlev07} Finally, we argue
that the interplay of long-range disorder with localized (magnetic) states in
graphene
offers a scenario that conciliates the experimental findings of
Refs.~\onlinecite{Chen11,Nair12,McCreary12} regarding the Kondo effect.


{\it Model Hamiltonian.}
At low concentrations, vacancies give rise to quasi-localized midgap states.
\cite{Pereira06,Pereira08} Since the latter are orthogonal to the conduction
band $\pi$-like states, there is no hybridization and, hence, no mechanism
allowing for Kondo physics. Recently, vacancy reconstructions with Jahn-Teller
out-of-plane lattice distortions have been put forward as a coupling mechanism
between localized and conduction band states. \cite{Cazalilla12,Kanao12,
Mitchell:Phys.Rev.B:75104:2013} 
The resulting effective model involves the coupling of the localized level with
a $\pi$-character conduction band with a log-divergent hybridization function,
\cite{Cazalilla12} whose rich phase diagram has been studied with NRG.
\cite{Mitchell:Phys.Rev.B:75104:2013} However, the special lattice
reconstruction around the vacancy on which the model relies is not supported by
most {\it ab initio} calculations.\cite{Palacios12,Nanda12,Casartelli13} Also,
this model predicts a large suppression of $T_K$ at small doping
\cite{Cazalilla12} that is at odds with the experiment. \cite{Chen11}

We follow an alternative route and investigate the effects of disorder, other
than vacancies, ubiquitous in graphene samples.\cite{Mucciolo10} For
simplicity, we consider only long range disorder due, for instance, to charge
puddles or to charges trapped at the substrate. In this way, we avoid
mechanisms that can give rise to additional localized states, potentially
obscuring our analysis.

The nearest neighbor tight-binding Hamiltonian for a monolayer
graphene sheet with a single vacancy reads
\begin{equation}
\label{eq:Hv}
H_{\rm v} = - t \sum_{\langle i, j \rangle} |i\rangle \langle j| +
t \sum_{\langle {\rm v}, i \rangle} | {\rm v}\rangle \langle i|  + \mbox{H.c.},
\end{equation}
where $\langle \cdots \rangle$ indicates a sum over nearest-neighbor atomic
sites and $t$ is the hopping term. The second term at the right-hand side of
Eq.~\eqref{eq:Hv} decouples the site v from the honeycomb lattice. We remove
the latter state from the Hilbert space, mimicking a vacancy.

The solution of $H_{\rm v} | \phi \rangle = \ve_\phi  | \phi\rangle$ gives
extended states with non-zero energy $\{ |\nu \rangle \}$ and a single
zero-energy quasi-localized state $|0\rangle$.\cite{Pereira08} The wave
function $\langle {\bm r}|0\rangle$ oscillates on the scale of the lattice
parameter $a$ and decays with the inverse distance to the vacancy.
\cite{Pereira06,Nanda12}

We introduce disorder by adding $U_{\rm dis} = \sum_{i \neq {\rm v}}| i \rangle
U_i \langle i |$ to our model Hamiltonian. $U_i=U_{\rm dis}({\bm r}_i)$ is the
local potential at the $i$th site for a given disorder realization. For
simplicity, we consider $U_{\rm dis}$ to be a Gaussian correlated random local
potential, namely, $\left\langle U_{\rm dis}(\bm{r})U_{\rm
dis}(\bm{r}^\prime)\right\rangle =\pi\xi^2(\delta W)^{2}(N_{\rm imp}/{\cal A})
\exp(-|\bm{r-\bm r}^{\prime}|^{2}/4\xi^2)$,
characterized by $N_{\rm imp}/{\cal A}$, $\delta W$ and $\xi$, the density of
scattering centers per unit area, disorder potential strength and range,
respectively. We take $\xi$ larger than the lattice parameter to ensure long
range disorder.

To single out the $|0\rangle$ state and to explicitly describe its coupling to the extended
states that form the conduction band, we introduce the projection operators $P = \sum_\nu
|\nu \rangle\langle \nu |$ and $Q = |0\rangle\langle 0 |$, with $P+Q=1$.

The single-particle model Hamiltonian $H= H_{\rm v} + U_{\rm dis}$ is
written as $H =  H_{PP}+H_{PQ}+H_{QP}+H_{QQ}$.
The projection into the localized state reads
\be
H_{QQ}= |0 \rangle \langle 0  | (H_{\rm v} + U_{\rm dis}) |0 \rangle \langle 0 |
= |0 \rangle \ve_{0}^{\rm dis} \langle 0 |,
\ee
where $\ve_{0}^{\rm dis} =  \langle 0  | U_{\rm dis} |0 \rangle$.  The energy shift of the
localized state,  $\ve_0^{\rm dis}$, scales with $\delta W$ and can be either positive or negative,
depending on the disorder realization potential.
The coupling term is written as
$H_{PQ} = \sum_\nu |\nu \rangle \langle \nu  | U_{\rm dis} |0 \rangle \langle 0|$,
since $H_{\rm v}| 0 \rangle=0$.
The projection into extended states reads
\be
H_{PP} = \sum_\nu   |\nu \rangle \ve_\nu \langle \nu | + \sum_{\nu,\nu'} |\nu
\rangle \langle \nu  | U_{\rm dis} |\nu'  \rangle \langle \nu'|.
\ee

In general $ \langle \nu  | U_{\rm dis} |\nu'  \rangle\neq 0$. Hence, it is
convenient to diagonalize $H_{PP}$ as $H_{PP} | \beta \rangle = \ve_\beta
|\beta\rangle$ and  write the Hamiltonian $H$ in the $\{ |\beta\rangle \}$
basis. For that purpose we introduce the projection operator $P' = \sum_\beta |
\beta \rangle \langle \beta |$ and write the single-particle model Hamiltonian
as
\be
H = H_{P'P'} + H_{P'Q} + H_{QP'} + H_{QQ}.
\label{eq:ProjectedH}
\ee
While $H_{QQ}$ remains unchanged, the projection of $H$ into the extended
states is now diagonal by construction. The modified coupling term reads
\be
H_{P'Q} = \sum_\beta |\beta \rangle \langle \beta  | U_{\rm dis} |0 \rangle \langle 0| \equiv
\sum_\beta |\beta \rangle t_{\beta 0} \langle 0 |,
\label{eq:CouplingTerm}
\ee
showing that long-range disorder provides a natural coupling mechanism
between extended and localized states.

We use the tight-binding orbitals and site amplitudes $\langle i|0\rangle$ to
calculate the Coulomb energy $U$ for double occupation of the midgap state. We
find that $U$ scales with system size as $(\log L)^{-2}$, in agreement with
scaling arguments using an envelope function approximation for $\langle
i|0\rangle$.\cite{Cazalilla12} The standard literature values, $U_{\rm
local}/t\sim 3.5$, for the graphene on-site Coulomb interaction
\cite{Schuller13} lead to a $U$ of the order of eV for a graphene sheet of $L
\sim 1\,\mu$m on SiO$_2$, an estimate significantly larger than that of
Ref.~\onlinecite{Cazalilla12}. We stress that our model considers a single
vacancy. For a realistic case of diluted vacancies, the midgap states become
more localized and $U$ increases.

In summary, our model consists of a disordered Hamiltonian, $H_{\rm
v}+U_{dis}$, plus an interaction term to account for a double occupancy of the
vacancy-generated state. The resulting Hamiltonian can be mapped into an
Anderson-like model of a localized state coupled to a continuous band with an
energy dependent density of states $\rho_{\rm dis}(\omega)$. We define
$\omega=\epsilon-\mu(V_g)$, the energy relative to the Fermi level. The energy
$\omega$ varies within the range $-D-\Delta \mu \leq \omega \leq D- \Delta
\mu$, where $D$ is the half-bandwidth and $\Delta \mu = \mu(V_g)-\mu(0)$ is the
Fermi energy relative to its value at the charge neutrality point $\mu(0)$.

In second quantization, the model Hamiltonian $H_A$ is cast as
$H_{A}=H_{\state}+H_{\band}+H_{s-b}$, namely,
\begin{align}
\label{Eq:HamiltonianTerms}
H_{\state} = &\, \delta\varepsilon \; n_{{\rm 0} \sigma} + U n_{{\rm 0} \uparrow}n_{{\rm 0} \downarrow}\nonumber\\
H_{\band} = &\, \int^{D-\Delta\mu}_{-D-\Delta \mu} d \omega \; \omega \;
c^{\dagger}_{\omega \sigma}
c^{}_{\omega \sigma} \\
H_{s-b} = &\,  \int^{D-\Delta \mu}_{-D-\Delta \mu} d \omega \sqrt{\frac{\Gamma_{\rm dis}(\omega)}{\pi}}
\left( c^{\dagger}_{{\rm 0} \sigma} c^{}_{\omega \sigma}+\mbox{H.c.} \right)\; ,\nonumber
\end{align}
where $\delta\varepsilon=\ve_{0}^{\rm dis}-\mu(V_g)$ is the midgap state energy
relative to the Fermi level. The remaining notation is standard:
$c^{\dagger}_{{\rm 0}\sigma}$ ($c^{}_{{\rm 0} \sigma}$) creates (annihilates)
an electron with spin $\sigma$ at the localized state and $n_{{\rm 0}
\sigma}=c^{\dagger}_{{\rm 0} \sigma}c^{}_{{\rm 0} \sigma}$ is the number
operator. The electron band states $\beta$ are treated in the energy
representation. Accordingly,  $c^{\dagger}_{\omega \sigma}$ ($c^{}_{\omega
\sigma}$) creates (annihilates) an electron with spin $\sigma$ and energy
$\omega$ in the (disordered) graphene band. The coupling between the band and
the localized state is written in terms of the hybridization function
$\Gamma_{\rm dis}(\omega)=\pi \sum_{\beta} |t_{\beta {\rm 0}}|^2
\delta(\omega-\epsilon_{\beta})$. The latter is a key element in NRG
logarithmic discretization of the conduction band.\cite{Gonzalez-Buxton98}


{\it Results.} We study the model given by Eq.\ (\ref{Eq:HamiltonianTerms})
using Wilson's NRG method. \cite{KrishnamurthyWW80,Bulla08} We calculate
quantities that characterize the different phases of the system, such as the
occupation $\langle n_{\rm 0} \rangle (T)$ and the impurity magnetic moment
$m^2(T)\equiv T\chi_{\rm imp}(T)/(g \mu_B)^2$, where $\chi_{\rm imp}$ is the
``impurity" (localized state) contribution to the magnetic susceptibility.
\cite{KrishnamurthyWW80,Bulla08}

Before addressing disorder effects, it is instructive to discuss a simpler case. Let us consider
 $\rho(\omega)=\rho_0|\omega-\Delta\mu|/D$, the density of states of pristine graphene,
and  $\Gamma_{\rm dis}(\omega)= \Gamma_0|\omega-\Delta\mu|/D$, where $\Gamma_0$
is chosen as the hybridization energy scale at the band edge.
\cite{Gonzalez-Buxton98} This toy-model parametrization of $\Gamma_{\rm
dis}(\omega)$ is rather naive, but serves the purpose of  guiding the
discussion. We improve it below, when we address a realistic disorder model.
Here, disorder manifests itself mainly by shifting $\varepsilon_0^{\rm dis}$.

For $\Delta\mu=0$, corresponding to the charge neutrality point, the density of
states vanishes as a power law $\rho(\omega)\sim|\omega|^r$. Quantum impurity
models that display such feature are generically referred to as ``pseudogapped
models"
\cite{Withoff90,Gonzalez-Buxton98,Fritz:Phys.Rev.B:214427:2004,Silva:096603:2006,Vojta10}
and present interesting properties such as a quantum phase transition (QPT) for
a critical set of model parameters. For the pseudogap Anderson model with
$r\!=\!1$, the QPT occurs for particle-hole asymmetric situations and is
characterized by a (unstable) fixed point with ``valence fluctuation"
properties: $m^2(T\!\rightarrow\!0)\!=\!1/6$ and $\langle n_{\rm 0} \rangle
(T\!\rightarrow\!0)\!=\!2/3$.
\cite{KrishnamurthyWW80,Fritz:Phys.Rev.B:214427:2004} A quantum phase
transition occurs at $\delta \epsilon = \delta \epsilon_c$, separating
``empty-orbital" ($\langle n_0 \rangle (T\!\rightarrow\!0) \sim 0$ and
$m^2(T\!\rightarrow\!0) \sim 0$ for $\delta \epsilon > \delta \epsilon_c$) and
local-moment ($\langle n_0 \rangle(T\!\rightarrow\!0) \sim 1$ and
$m^2(T\!\rightarrow\!0) \sim 1/4$ for $\delta \epsilon < \delta \epsilon_c$)
phases.\cite{SuppMaterial}

This behavior is markedly different from that described by the usual
($r\!=\!0$) Anderson impurity model. In the latter, the band is metallic,
leading to Kondo screening of the impurity magnetic moment for $\delta
\epsilon$ in the range $-U < \delta \epsilon < 0$. For $-U < \delta \epsilon
\ll 0$, the crossover to the Kondo regime is characterized by $\langle n_{\rm
0} \rangle(T\!\rightarrow\!0) \sim 1$ and $m^2(T\!\rightarrow\!0)\rightarrow
0$. The crossover energy scale is the Kondo temperature $T_K$. For $\delta
\epsilon
> 0$ and $\langle n_{\rm 0} \rangle(T\!\rightarrow\!0)\!\sim\!0$, the system
enters a different regime, characterized by an ``empty-level" or
``frozen-impurity" fixed point \cite{KrishnamurthyWW80} without Kondo
screening, although $m^2(T)\rightarrow 0$ for $T\rightarrow 0$. The transition
to the empty-level fixed point is associated with a crossover scale
$T^*\!\gg\!T_K$ of the order of $\Gamma_0$. \cite{KrishnamurthyWW80_1}

Long-range disorder changes this picture dramatically. Our microscopic disorder
model gives rise to realization-dependent fluctuations in $\varepsilon_0^{\rm
dis}$ and $\Gamma_{\rm dis} (\omega)$. It describes the low-energy physics of
the system in terms of a \textit{disordered} effective Anderson model.
\cite{Miranda:Phys.Rev.Lett.:264:2001,Mirandareview,
Cornaglia:Phys.Rev.Lett.:117209:2006,Zhuravlev07} For any given disorder
realization, the NRG analysis of $H_A$, Eq. \eqref{Eq:HamiltonianTerms}
requires $\varepsilon_0^{\rm dis}$ and $\Gamma_{\rm dis}(\omega)$ as an input.
To this end, we proceed as follows.

We obtain the density of states by an exact diagonalization of  the
single-particle Hamiltonian $H$ in a periodic honeycomb lattice of $N_{\rm s}$
sites, with a vacancy site at its center. We take $N_{\rm s}\gg 1$ and
approximate the continuum by the spectrum calculated at the superlattice
$\Gamma$ point ($k=0$).  We smoothen $\rho_{\rm dis}(\omega) = \sum_{\beta}
\delta(\omega-\epsilon_{\beta})$ by making $\rho_{\rm dis}(\omega) \approx
N(\omega)/\Delta E$, where $N(\omega)$ is the number of band states in the
energy window $\omega-\Delta E/2$ and $\omega+\Delta E/2$. Since $N_{\rm s}$ is
finite, the spectrum of $\epsilon_{\beta}$ has a small gap at low energies.
Therefore, the choice of $\Delta E$ is a compromise between the enhancement of
the fluctuations due to disorder and the smearing of the finite-size gap. The
same procedure is used to compute the effective energy-dependent coupling
$|t(\omega)|^2$. We define $|t(\omega)|^2$ as the average of $|t_{\beta {\rm
0}}|^2$ in the window $\omega -\Delta E/2 \leq \epsilon_{\beta}\leq \omega +
\Delta E/2$. The hybridization function is approximated as $\Gamma_{\rm
dis}(\omega) \approx \pi |t(\omega)|^2 \rho_{\rm dis}(\omega)$.

\begin{figure}[t]
\begin{center}
\includegraphics[width=1.0\columnwidth]{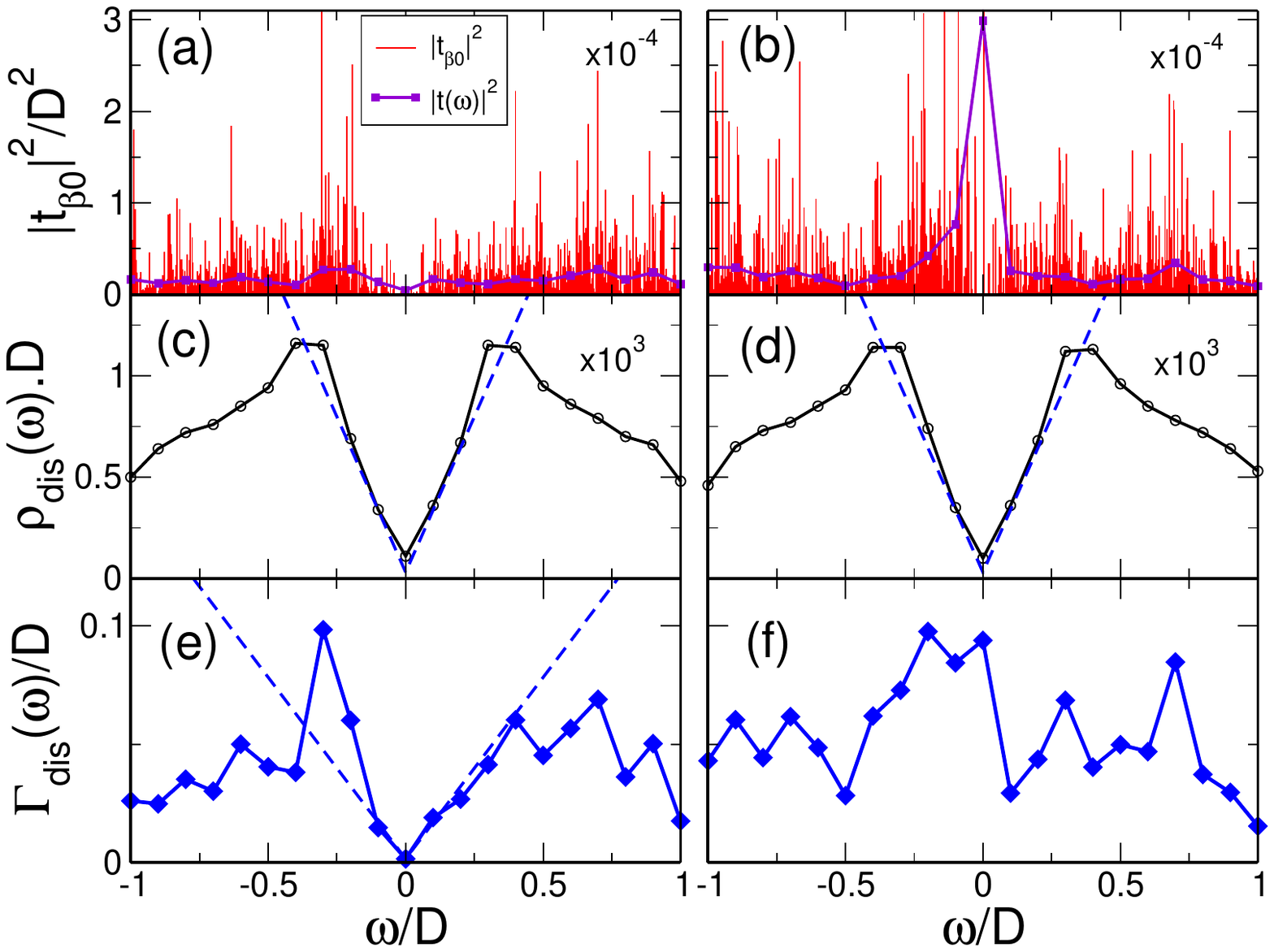}
\caption{(Color online) Examples of disorder realizations leading to
``pseudogap" (left panels) and ``metallic" (right panels) behavior. Raw data
for $|t_{\beta {\rm 0}}|^2$ and the corresponding energy-averaged function
$|t(\omega)|^2$ (a--b). Panels (c--d) show the  density of states $\rho_{\rm
dis}(\omega)$, which is linear for $|\omega|/D \ll 1$ in both cases. The
hybridization function $\Gamma_{\rm dis}(\omega)$ is also linear in the
pseudogap case (e) in contrast with the ``metallic" case (f). The latter
behavior stems from strong fluctuations in the coupling between the localized
state $|0\rangle$ and states $|\beta\rangle$ with energies $|\ve_\beta|$ close
to the Fermi energy, shown in (b).} \label{fig:t_E}
\end{center}
\end{figure}

In Fig.~\ref{fig:t_E} we show $|t_{\beta {\rm 0}}|^2$ for two disorder
realizations for a disorder strength $\delta W=0.316t$, range $\xi = 3 a$,
system size $N_{\rm s}= 40 \times 40$ and $N_{\rm imp}=N_{\rm s}/10$. The
results show that $|t(\omega)|^2$ [Figs.\ \ref{fig:t_E}(a)
and~\ref{fig:t_E}(b)] is essentially independent of energy for large
$|\omega|$. Furthermore, we note that $|t(\omega)|^2$ becomes increasingly
sensitive to fluctuations as $|\omega|$ becomes smaller (the region with
interest to Kondo physics), since the number of states in this energy range is
relatively small. This can lead to rather strong fluctuations in $|t(\omega\
\sim\ 0)|^2$. Thus, although the density of states retain, in general, the
characteristic linear behavior near the charge neutrality point
[Figs.~\ref{fig:t_E}(c) and~\ref{fig:t_E}(d)], fluctuations in $|t(\omega \sim
0)|^2$ lead to a
 ``metallic" character (in the NRG sense)
near the Fermi energy in $\Gamma_{\rm dis}(\omega)$, as shown in
Fig.~\ref{fig:t_E}(f).

For every disorder realization we use the procedure described above to compute
$\Gamma_{\rm dis}(\omega)$ [Figs.~\ref{fig:t_E}(e) and~\ref{fig:t_E}(f)]. The
latter and $\ve_0^{\rm dis}$ are used as inputs to the NRG calculations. We
note that different choices of $\Delta E$ do not appreciably alter the
low-energy part of $\Gamma_{\rm dis}(\omega)$ as long as $\Delta E$ is of the
order of the finite-size-induced gap.
The Kondo temperature $T_K$ and $T^*$ are obtained from the analysis of the
behavior of the magnetic moment $m^2(T)$ and the occupation $\langle n_{\rm 0}
\rangle (T)$ versus $T$.\cite{SuppMaterial}

\begin{figure}[t]
\begin{center}
\includegraphics[width=1.0\columnwidth]{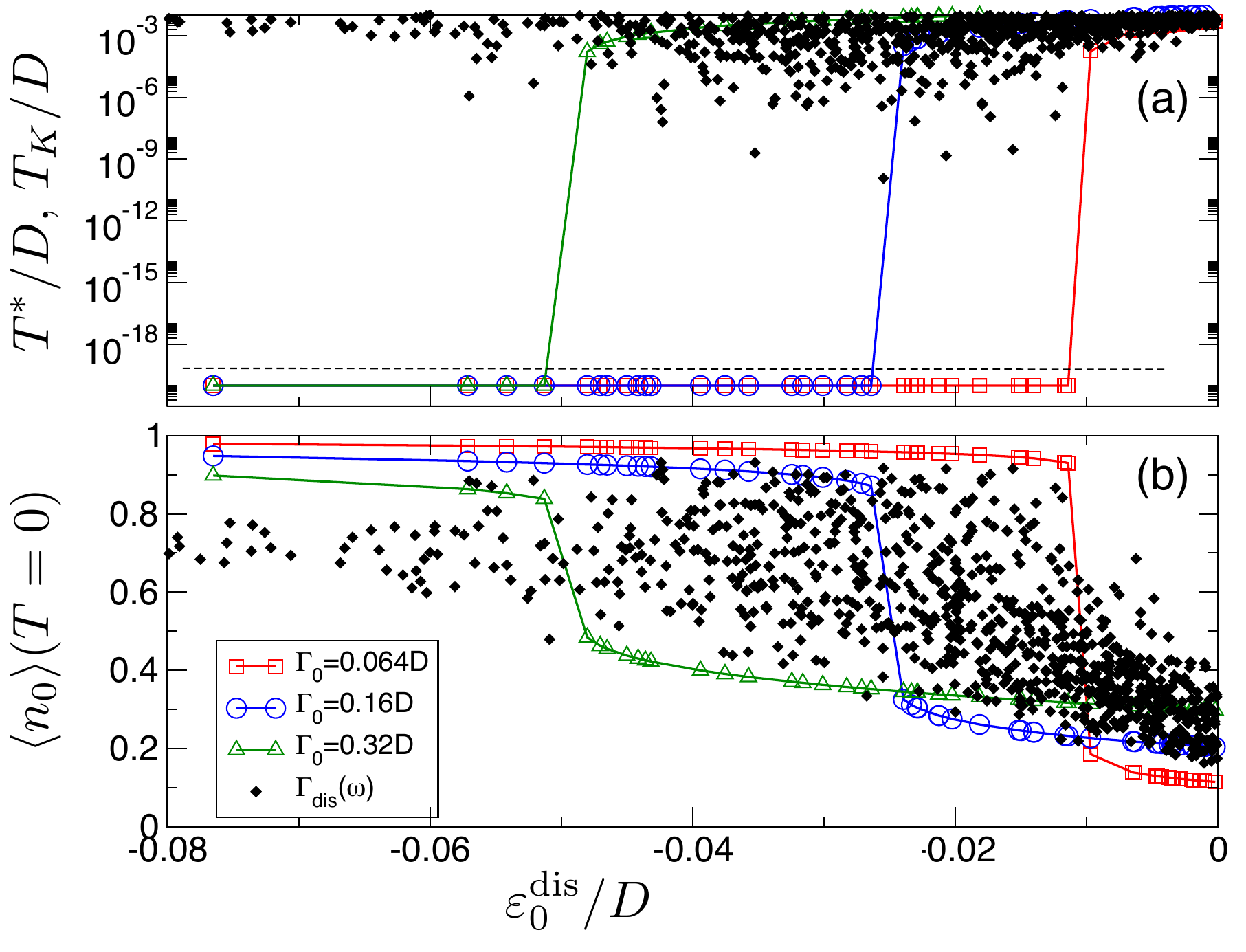}
\caption{(Color online) Crossover ($T^{*}$) and Kondo ($T_K$) temperatures (a) and occupation
$\langle n_0 \rangle$ (b) versus $\ve_{0}^{\rm dis}$ obtained for different disorder realizations
(diamonds) and for the pseudogap
toy model, $\Gamma(\omega)=\Gamma_0 | \omega|$
(open symbols and solid lines).}
\label{fig:TKvsedis}
\end{center}
\end{figure}

Figure \ref{fig:TKvsedis} shows the NRG results for $T_K$ (or $T^*$)
 and $\langle n_0 \rangle$ for 10$^3$ disorder realizations.
The single-particle parameters are the same as in Fig.~\ref{fig:t_E}, $U=0.5D$,
and the system is at the charge neutrality point, $\Delta \mu=0$. We have
stopped the NRG calculations at scales of the order $10^{-20}D$ so this scale
(dashed line) defines the ``zero temperature."

To contrast with Kondo pseudogap physics, Fig.~\ref{fig:TKvsedis} also shows
results for the pseudogap toy model $\Gamma(\omega)=\Gamma_0 | \omega|$, for
different values of $\Gamma_0$. In this case (open symbols),  both $T^*$  and
$\langle n_0 \rangle$ versus the disorder dependent $\ve^{\rm dis}_0$  show
``jumps," marking the well-known quantum phase transitions
\cite{Gonzalez-Buxton98} of the linear pseudogap Anderson model: For a fixed
$U$, they occur at critical values of the impurity level energy $\ve_{0}^{\rm
dis}=\ve_{0}^{*}(\Gamma_0)$ separating empty-orbital ($\ve_{0}^{\rm
dis}>\ve_{0}^{*}$) and local-moment ($\ve_{0}^{\rm dis}<\ve_{0}^{*}$) phases.
The latter is characterized by vanishing $T^*$ and $\langle n_0 \rangle
\rightarrow 1$, while the former has non-zero $T^*$ and $\langle n_0 \rangle
\rightarrow 0$.\cite{SuppMaterial}

The long-range disordered model (diamonds) shows important differences:
Fluctuations in the disorder potential lead to Kondo ground states,
characterized by $\langle n_0 \rangle \rightarrow 0.8-1.0$ with a non-vanishing
$T_K$. A striking consequence is that the sharp features of pseudogap-related
quantum phase transitions at $\Delta\mu=0$  are no longer evident. Instead, the
disorder-induced filling of the pseudogap leads to the formation of Kondo
singlets, which dominate the low temperature properties.

\begin{figure}[t]
\begin{center}
\includegraphics[width=1.0\columnwidth]{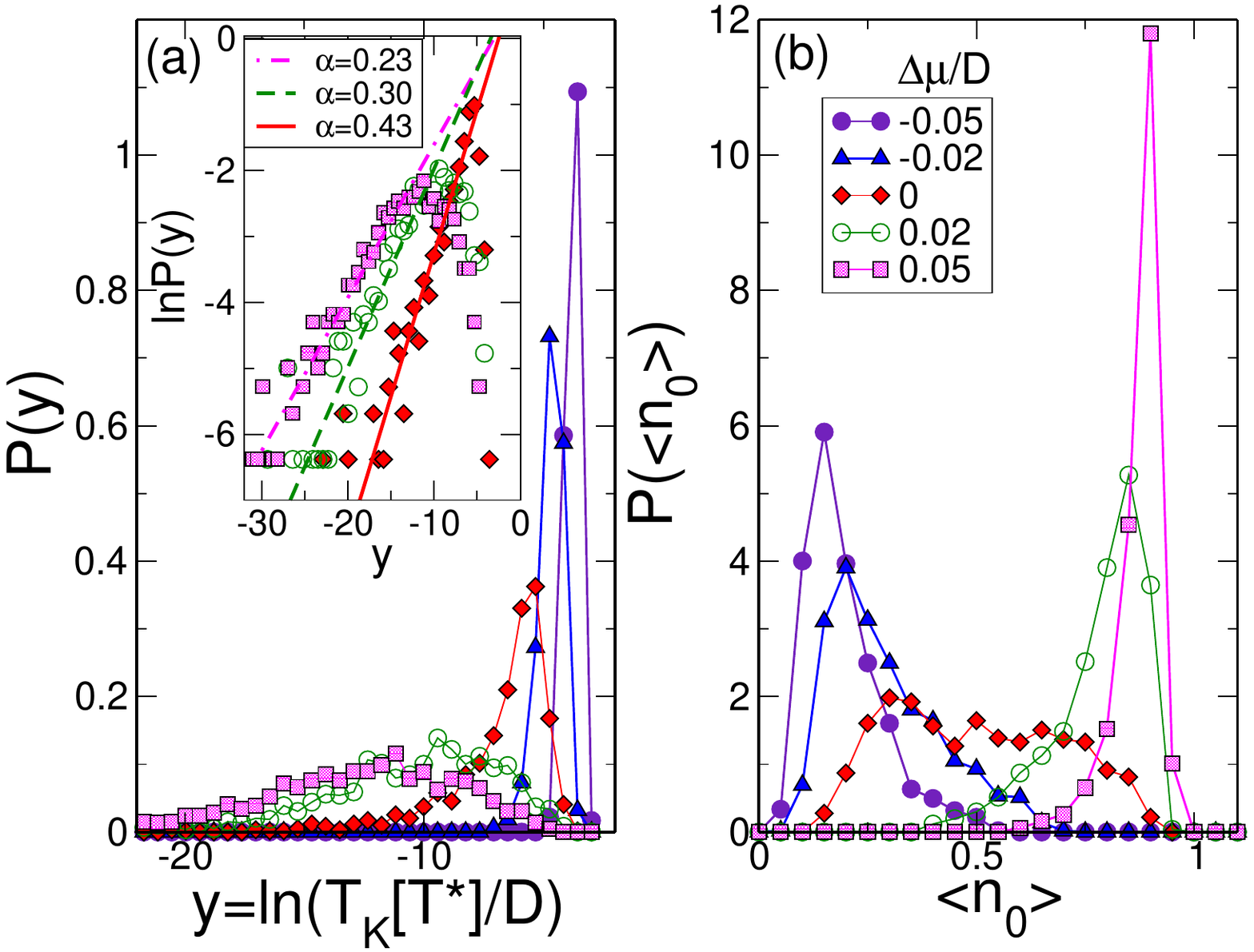}
\caption{(Color online) Normalized distributions of the crossover/Kondo temperatures (a)
and occupations (b) for different values of the chemical potential $\Delta \mu$.
Inset: Power-law fitting of the low-$T_K$ tails of the distributions for $\Delta \mu \geq 0$.}
\label{fig:HistTK}
\end{center}
\end{figure}

The picture that emerges is that the vacancy induced state behaves as an
Anderson impurity embedded in a ``disordered metal"
\cite{Zarand96,Lewenkopf05,Miranda:Phys.Rev.Lett.:264:2001,
Mirandareview,Cornaglia:Phys.Rev.Lett.:117209:2006,Zhuravlev07}, with
realization dependent model parameters. The disorder fluctuations give rise to
a distribution of Kondo temperatures. Figure \ref{fig:HistTK} shows the
distributions $P[\log(T^*)]$ (or $P[\log(T_K)]$ in the Kondo regime) and
$P\left(\langle n_0 \rangle \right)$ for $\sim 10^3$ disorder realizations and
different values of $\Delta\mu$.

For  large $\Delta \mu<0$, there is a predominance of positive values of
$\delta \varepsilon= \ve_0^{\rm dis} - \Delta \mu$, favoring small occupations
[Fig.~\ref{fig:HistTK}(b)] and relatively large crossover temperatures $T^*$
[Fig.~ \ref{fig:HistTK}(a)]. This behavior is very clear for $\Delta \mu
=-0.05D$, but changes qualitatively as $\Delta \mu$ increases. Already at
$\Delta\mu=-0.02D$, distinct ``tails" in the distributions of $\langle n_0
\rangle$ and $\log(T^*)$ can be seen.

At $\Delta\mu=0$, the distributions reflect the trends shown in
Fig.~\ref{fig:TKvsedis}, with $P[\log(T^*)]$ displaying two clear features: a
sharp peak at larger values of $T^*$ and a long log-distributed tail. The
realizations contributing to the peak in $P[\log(T^*)]$ lead to small values of
$\langle n_0 \rangle$, which correspond to the ``tail" in $P\!\left(\langle n_0
\rangle \right)$ shown in Fig.~\ref{fig:HistTK}(b).

For  $\Delta\mu>0$, the disordered Kondo phase clearly dominates, characterized
by $P(\log(T_K))$ with long logarithmic tails along with a sharp peak in
$P\left(\langle n_0 \rangle \right)$ around $\langle n_0 \rangle \sim 1$. A
more careful analysis (inset in Fig.~\ref{fig:HistTK}) shows that, for small
$T_K$, the Kondo temperature distributions follow a power-law behavior
$P(T_K)\propto T^{(\alpha-1)}_K$ with $\alpha\sim0.2-0.5$, depending weakly on
$\Delta \mu$. Such behavior has been previously found in disordered Anderson
systems,\cite{Miranda:Phys.Rev.Lett.:264:2001,Mirandareview} where the
interpretation for the divergent behavior of $P(T_K)$ for small $T_K$ with
non-universal exponents was given in terms of a quantum Griffiths phase and
disorder-induced non-Fermi-liquid behavior.
\cite{Neto:PhysicalReviewLetters:3531:1998,Mirandareview}

The exponent $\alpha$ is known to depend on the disorder strength and one
expects $\alpha\!<\!1$ and divergent behavior in $P(T_K)$ only for strong
disorder.\cite{Miranda:Phys.Rev.Lett.:264:2001,Mirandareview} Interestingly, in
Fig.~\ref{fig:HistTK}, the disorder strength was kept fixed giving
$\alpha\sim0.2-0.5$, with a weak dependency with $\Delta \mu$. This feature is
a consequence of the increased broadening of the Kondo temperature
distributions, shown in Fig.~\ref{fig:HistTK}-a. As the system enters deeper in
the Kondo regime (increasing $\Delta \mu$), small fluctuations in the
single-particle parameters produces large fluctuations in the Kondo scale.
\cite{Mirandareview} This leads to longer and flatter logarithmic tails, with
smaller values of $\alpha$.
%


{\it Conclusions.} The effect of disorder in our system is twofold. First, it
provides a simple mechanism, so far overlooked, to couple the localized state
with the graphene band. Secondly, disorder fluctuations lead to a distribution
of Kondo temperatures $P(T_K)$ with a power-law divergence at low $T_K$. This
is  consistent with the presence of a Griffiths phase and allows for the
interesting possibility of detecting disorder-induced non-Fermi-liquid behavior
\cite{Neto:PhysicalReviewLetters:3531:1998} in transport experiments in
graphene.

Assuming a very dilute vacancy concentration and that the resistivity is
dominated by the localized states with the largest $T_K$, our simulations are
consistent with experimentally measured $T_K$ of the order of a few Kelvin
\cite{Chen11} with a weak dependence on $|\mu|$ as long as it stays close to
the charge neutrality point.  We find that the mean $T_K$ depends strongly on
the disorder strength. For less disordered samples where, for instance, charge
puddles fluctuations are smaller, $T_K$ would be  dramatically suppressed  and
one expects to observe only local magnetic moments. \cite{SuppMaterial} This
picture offers a unified scenario to interpret the puzzle posed by experiments.
\cite{Chen11,Nair12,McCreary12}

\textit{Note added:} Recently we became aware of new STM measurements of
Kondo-like resonances in Co adatoms on graphene deposited on a Ru$(0001)$
substrate.\cite{Ren:NanoLetters:4011:2014} The Kondo effect is attributed in
this case to an increase of rippling in the graphene sheet. These results are
consistent with the main argument we make in this paper: that long-range
disorder can play an important role for the observation of the Kondo effect in
graphene.


\textit{Acknowledgements:} We acknowledge helpful discussions with E.\
Mucciolo, E.\ Miranda, and K.\ Ingersent. This research is supported by
Brazilian funding agencies FAPERJ, FAPESP, and CNPq, and by the
INCT-–Nanomateriais de Carbono and PRP/USP-NAP Q--Nano initiatives.





\pagebreak \widetext
\begin{center}
\textbf{\large Supplemental Material for ``Disorder-mediated Kondo effect in
graphene"}
\end{center}
\setcounter{equation}{0} \setcounter{figure}{0} \setcounter{table}{0}
\setcounter{page}{1} \makeatletter



\section{Details of the NRG calculations}

Here we provide additional details on the NRG calculations performed in the
effective Anderson model (Eq.~(6) in the main paper). A key aspect is the
discretization of the effective graphene band which leads to an
energy-dependent hybridization function $\Gamma_{\rm dis}(\omega)$. To this
end, we have employed the standard
approach:\cite{Sup_Gonzalez-Buxton98,Sup_Bulla08} performing a logarithmic
discretization of a conduction band with arbitrary energy dependence through a
Lanczos procedure to map the continuous band to a one-dimensional Wilson chain.

For each disorder realization, we calculated standard thermodynamic quantities
in NRG: the impurity contribution to the susceptibility $\chi_{\rm imp}(T)$ or,
more precisely, the effective impurity magnetic moment $m^2(T)\equiv T\chi_{\rm
imp}(T)/(g \mu_B)^2$ (where $g$ and $\mu_B$ are the Land\'e g-factor and the
Bohr magneton, respectively) and the mean impurity occupation $\langle
n_0\rangle(T)$. In the calculations, we used charge and total SU(2) spin
quantum numbers retaining up to 1000 states, using the discretization parameter
$\Lambda=2.5$ and up to 100 sites in the Wilson chain, which corresponds to a
cut-off temperature $T_{\rm min} \approx 10^{-20}D$. Both $T_K$ and $T^*$ are
extracted from the NRG data by using the criteria $m^2(T_K)(m^2(T^*))=0.0701$.
\cite{Sup_KrishnamurthyWW80_1}

\subsection{Differences on ``pseudogap"  and ``metallic" regimes}

Here, for the sake of completeness, we discuss the main features (and
differences) of the NRG method applied to ``pseudogap"  and ``metallic" regimes
of the Anderson impurity model.

Using the procedure described above, NRG can be applied to Kondo or Anderson
``pseudogapped models"
\cite{Sup_Withoff90,Sup_Gonzalez-Buxton98,Sup_Fritz:Phys.Rev.B:214427:2004,Sup_Silva:096603:2006,Sup_Vojta10}
where the hybridization function can be written as
$\Gamma(\omega)=\Gamma_0|\omega|^r$ near the Fermi energy. It is instructive to
review some of the main features for the case $r\!=\!1$ and the differences to
the metallic case ($r\!=\!0$). The notation follows the one in the main paper.

\begin{figure}[h!]
\begin{center}
\includegraphics[width=0.7\columnwidth]{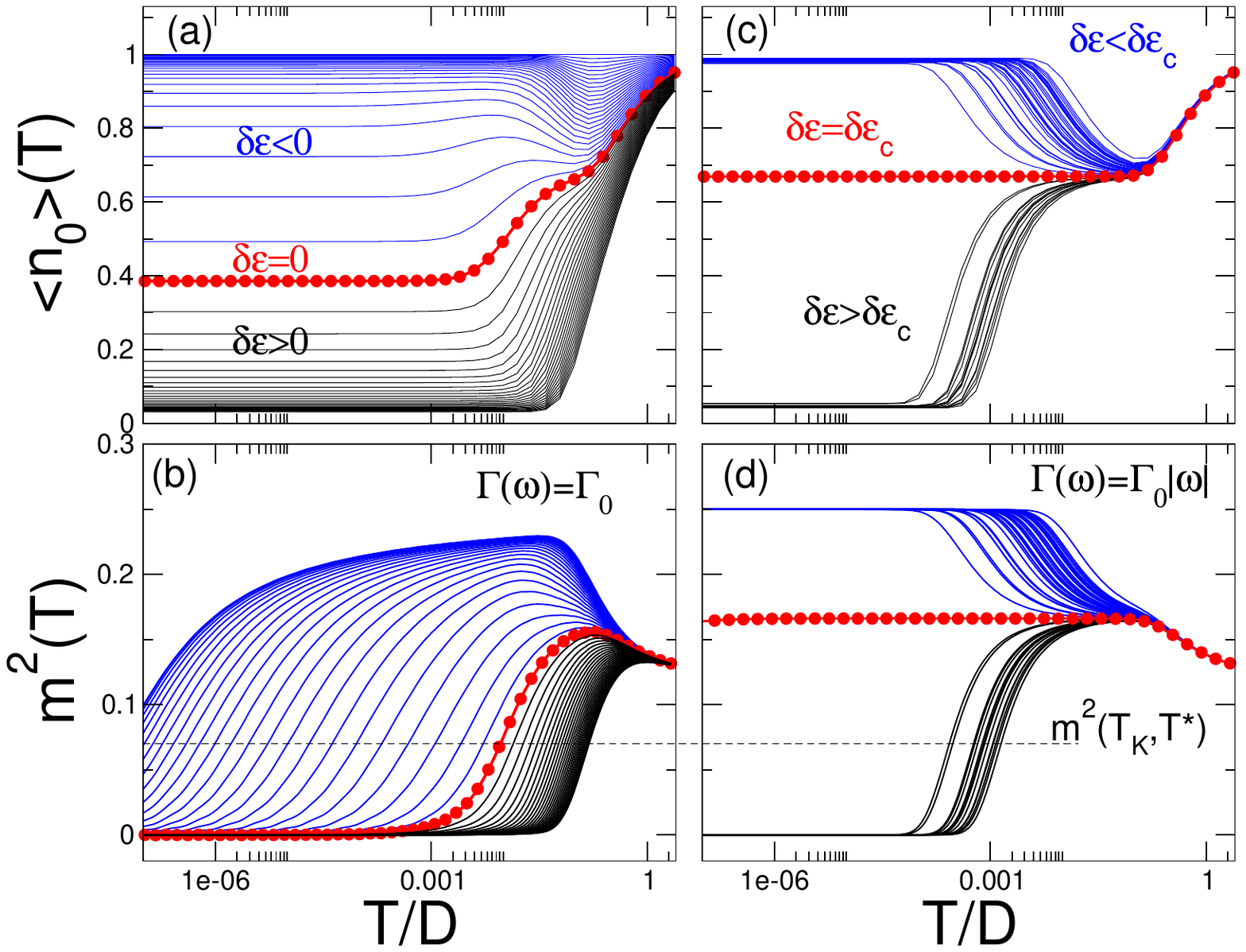}
\caption{ Level occupation $\langle n_0 \rangle (T)$ (a,c) and impurity
magnetic moment $m^2(T)\equiv T\chi_{\rm imp}(T)/(g \mu_B)^2$ (b,d) for the
metallic (left panels) and pseudogap (right) Anderson models with $-U \leq
\delta \epsilon \leq 0$ . In the pseudogap model, a quantum phase transition
(QPT) occurs at $\delta \epsilon = \delta \epsilon_c$ separating empty-orbital
($\langle n_{\rm 0} \rangle (T \rightarrow 0) \sim 0$ and $m^2(T \rightarrow 0)
\sim 0$ for $\delta \epsilon > \delta \epsilon_c$) and local-moment ($\langle
n_{\rm 0} \rangle (T \rightarrow 0) \sim 1$ and $m^2(T \rightarrow 0) \sim
\frac{1}{4}$ for $\delta \epsilon < \delta \epsilon_c$) phases. This is in
contrast with the smooth crossover from Kondo screening to the empty-orbital
regime in the metallic Anderson model (a,b). } \label{fig:NRG1}
\end{center}
\end{figure}

Figures \ref{fig:NRG1} (c-d) show typical  NRG results for the magnetic moment
$m^2(T)$ and occupation $\langle n_0\rangle(T)$ for the linear ($r\!=\!1$)
pseudogap Anderson model. Each curve corresponds to a different value of
$\delta \epsilon$, while keeping the chemical potential fixed at
$\mu(V_g)=\mu(0)$ (thus changing the impurity level $\ve_{0}^{\rm dis}$ only).
A quantum phase transition occurs at $\delta \epsilon = \delta \epsilon_c$,
separating ``empty-orbital" ($\langle n_{\rm 0} \rangle (T\!\rightarrow\!0)
\sim 0$ and $m^2(T\!\rightarrow\!0) \sim 0$ for $\delta \epsilon > \delta
\epsilon_c$) and local-moment ($\langle n_{\rm 0} \rangle(T\!\rightarrow\!0)
\sim 1$ and $m^2(T\!\rightarrow\!0) \sim \frac{1}{4}$ for $\delta \epsilon <
\delta \epsilon_c$) phases.

We note that this behavior is markedly different than the one expected for the
usual (non-pseudogap or $r\!=\!0$) Anderson impurity model
(Fig.~\ref{fig:NRG1}-a,b) in which no phase transition takes place. In such
case, the band is metallic ($\Gamma(\omega)=\Gamma_0$) and Kondo screening
occurs for $\delta \epsilon$ in the range $-U < \delta \epsilon \ll 0$.
The crossover to the Kondo regime is characterized by $\langle n_{\rm 0}
\rangle(T\!\rightarrow\!0) \sim 1$ and $m^2(T\!\rightarrow\!0)\rightarrow 0$.
The characteristic energy scale of the crossover is the Kondo temperature
$T_K$. For $\delta \epsilon \gg 0$, $\langle n_{\rm 0}
\rangle(T\!\rightarrow\!0)\!\sim\!0$, the system enters a different regime,
characterized by an ``empty-level" or ``frozen-impurity" fixed point
\cite{Sup_KrishnamurthyWW80_1} where no Kondo screening takes place even though
$m^ 2(T)\rightarrow 0$ for $T\rightarrow 0$. The transition to the empty-level
fixed point is characterized by a cross-over scale $T^*\!\gg\!T_K$, which is of
order of $\Gamma_0$.\cite{Sup_KrishnamurthyWW80_1}

\section{Influence of the disorder strength $\delta W$}

\begin{figure}[t]
\begin{center}
\includegraphics[width=0.7\columnwidth]{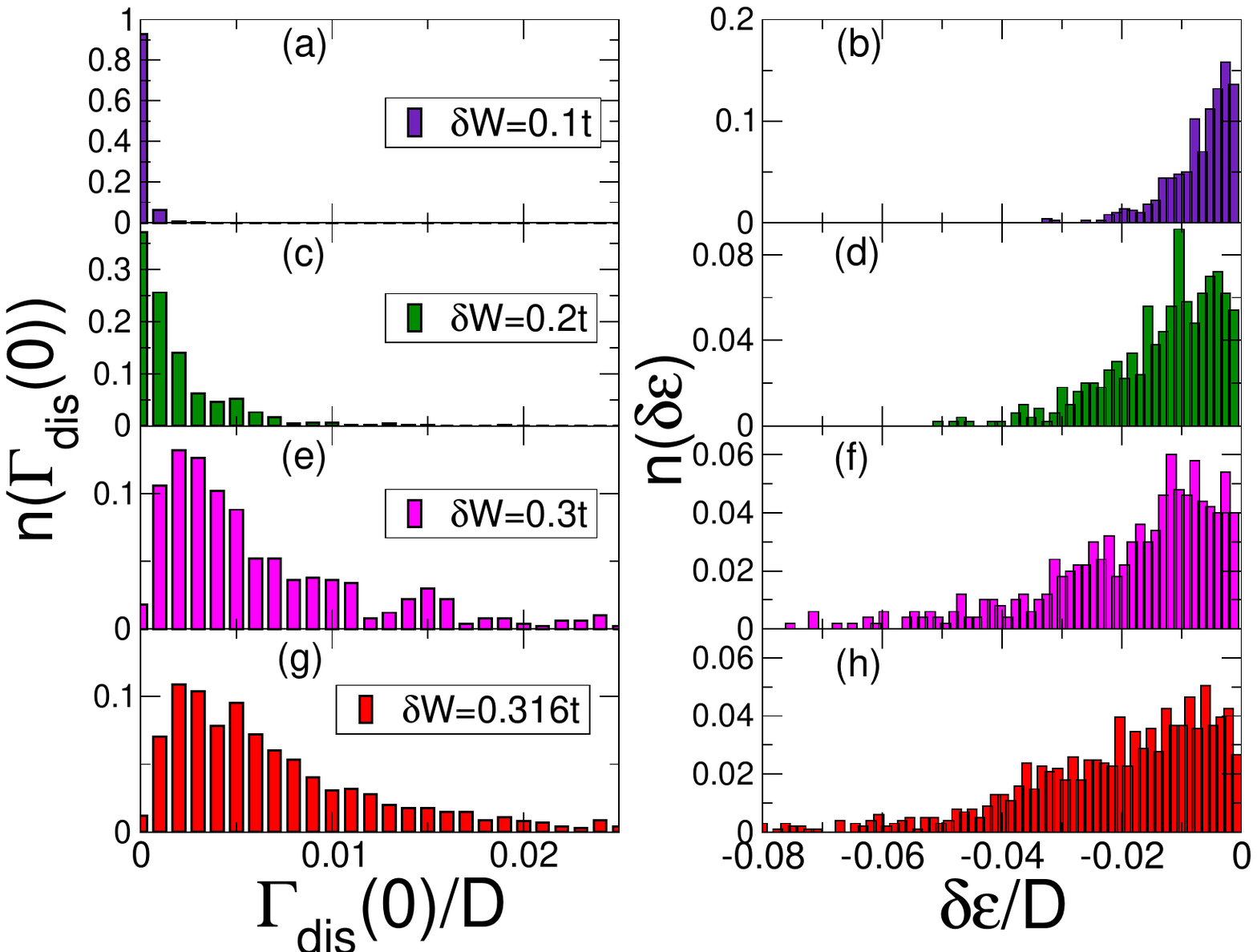}
\caption{(Color online) Distributions of $\Gamma_{\rm dis}(0)$ and
$\varepsilon_0^{\rm dis}$ for different disorder strengths $\delta W$. The
$\Gamma_{\rm dis}(0)$ distribution favors small values of $\Gamma_{\rm
dis}(0)$. Here we only consider distributions for $\varepsilon_0^{\rm dis}<0$,
the condition yielding to Kondo behavior. Notice that most of the cases remain
in the mixed valence regime $|\varepsilon_0^{\rm dis}| \ll U$. }
\label{fig:HistGamma0_diffW}
\end{center}
\end{figure}

In this section, we illustrate the effect of the disorder potential strength
$\delta W$ in the distributions shown in the main paper.

\subsection{Distributions of $\Gamma_{\rm dis}(0)$ and $\varepsilon_0^{\rm dis}$}

We begin with the realization-dependent parameters entering the effective
Anderson model (Eq. 6 in the main paper). Fig.~\ref{fig:HistGamma0_diffW} shows
the normalized histograms for 4 values of $\delta W$ over 500 realizations of
two key quantities: the hybridization function value at the Fermi energy
$\Gamma_{\rm dis}(\omega\!=\!0)$ (which will depend strongly in  the coupling
$|t_{\beta {\rm 0}}|^2$ between the localized state $|0\rangle$ and states
$|\beta\rangle$ with energies $|\ve_\beta|$ close to the Fermi energy) and the
midgap state energy $\delta \ve=\ve_{0}^{\rm dis}-\mu(V_g)$. For the results
shown in Fig.~\ref{fig:HistGamma0_diffW}, $\delta \ve=\ve_{0}^{\rm dis}$ since
we set $\mu(V_{g})=0$.

As $\delta W$ increases, both the width and the mean of these distributions
increase accordingly. This is expected since $\ve_{0}^{\rm dis}$ and $|t_{\beta
{\rm 0}}|^2$ are directly connected to the disorder potential. We notice that
even for the largest value used ($\delta W=0.316t$, used in the main paper),
the mean values of these distributions are over an order of magnitude smaller
than the local Coulomb repulsion $U$ ($0.5D$ in the calculations).

\subsection{Kondo temperature distributions}

\begin{figure}[t]
\begin{center}
\includegraphics[width=0.7\columnwidth]{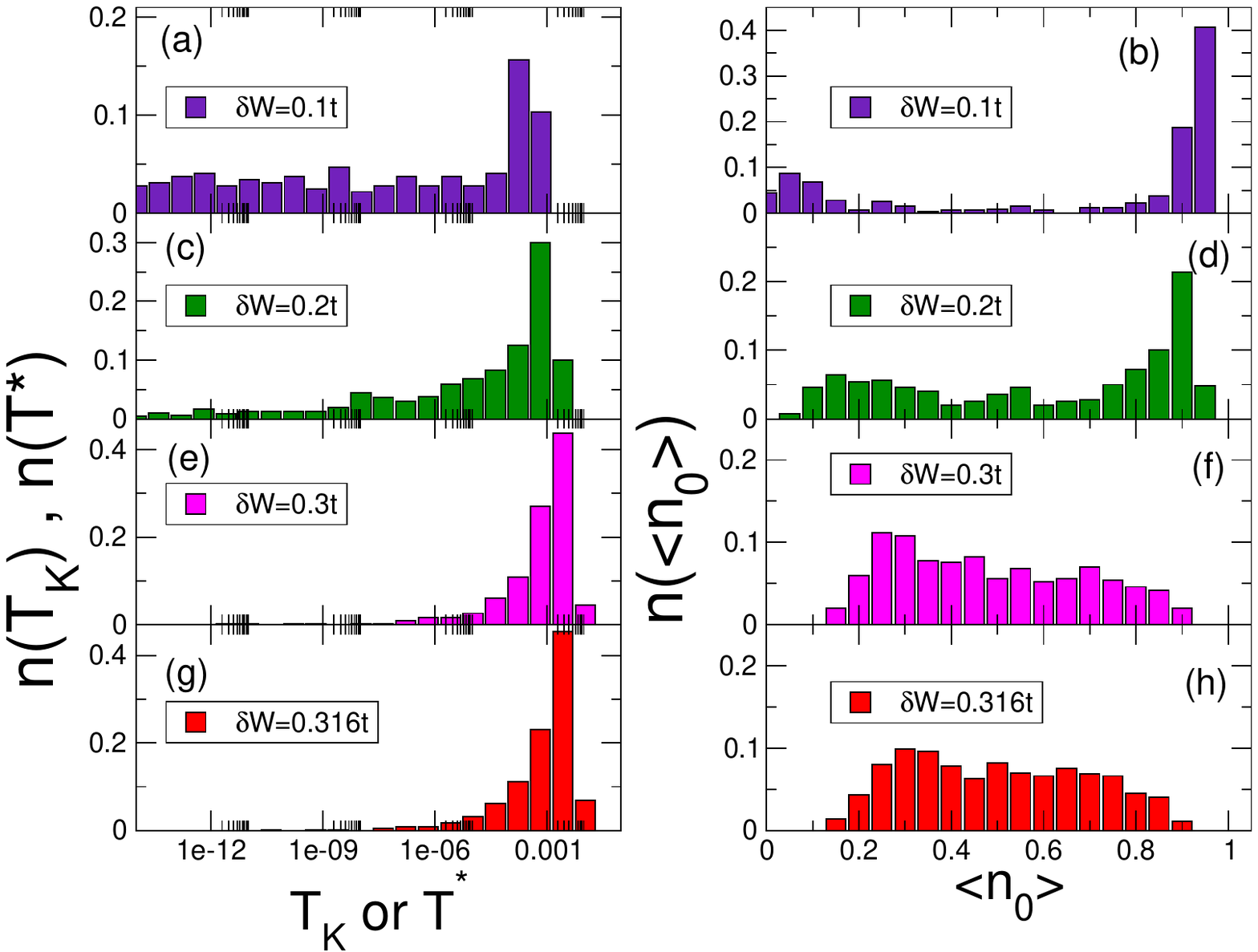}
\caption{(Color online) Kondo temperature (left panels) and impurity occupation
(right panels) distributions for different disorder strengths $\delta W/t$ and
$\Delta \mu=0$.}
\label{fig:HistTK_vsW}
\end{center}
\end{figure}

Following the procedure described in the main text, we use $\Gamma_{\rm
dis}(\omega)$  and $\delta \ve$ obtained for each individual realization as
input for the NRG calculations. Using the same realizations used to produce the
histograms shown in Fig.~\ref{fig:HistGamma0_diffW}, we obtain (normalized)
histograms for $T_K$ and $\langle n_0 \rangle$ shown in Fig.\
\ref{fig:HistTK_vsW} for $\Delta \mu=0$.

\begin{figure}[t]
\begin{center}
\includegraphics[width=0.7\columnwidth]{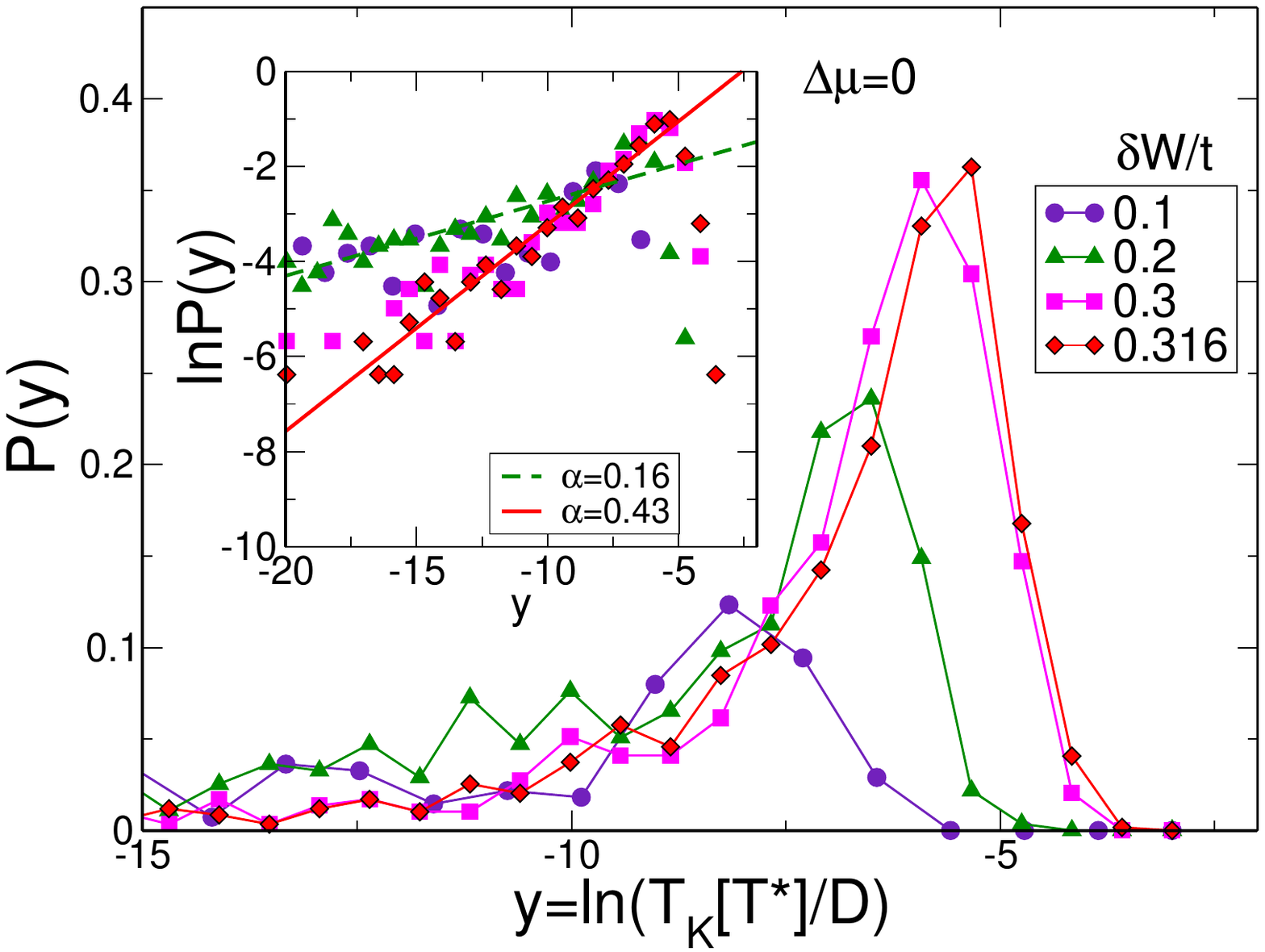}
\caption{(Color online) Kondo temperature distributions ($P(y)$ with
$y=\ln{(T_K)}$) for different disorder strengths $\delta W/t$ and $\Delta
\mu=0$. Inset: The disorder-dependent exponents $\alpha$ are obtained from the
tails of the distributions.}
\label{fig:HistlogTK_diffW}
\end{center}
\end{figure}

The $T_K$ histograms show the characteristic long logarithmic tail of the
Griffiths phase. The actual probability distributions are shown in Fig.~
\ref{fig:HistlogTK_diffW}. From the data, the exponents $\alpha$ can be
calculated. For $\Delta \mu=0$, the results show that $\alpha$ depends on
$\delta W$. In fact, $\alpha$ \textit{decreases} as the disorder strength
decreases. This is a consequence of the fact that disorder itself is the
mechanism inducing the Kondo effect in this system: the stronger the disorder,
the larger are the Kondo temperatures and the narrower is the $T_K$
distribution.

It should be noted that this trend was observed in a rather limited range of
$\delta W$. One of the reasons is that, numerically it becomes increasingly
difficult to obtain $\alpha$ in the low disorder limit: due to the smallness of
the typical couplings $\Gamma_{\rm dis}(0)$ and, more importantly, $|\delta
\ve|$, the typical $T_K$ values become exponentially small, rendering it
necessary to a substantial increase in the number of Wilson chain sites in the
NRG calculations.



\end{document}